# On the Humble Origins of the Brownian Entropic Force

Richard M. Neumann[a]
Five Colleges, Inc.
97 Spring Street
Amherst, Massachusetts 01002

**Abstract**

Recognition that certain forces arising from the averaging of the multiple impacts of a solute particle by the surrouding *solvent* particles undergoing random thermal motion can be of an entropic nature has led to the incorporation of these forces and their related entropies into theoretical protocols ranging from molecular-dynamics simulations to the modeling of quarkonium suppression in particle physics. Here we present a rigorous derivation of this Brownian entropic force by means of the classical Gibbs canonical partition function and in so doing provide a heuristic demonstration of its kinetic origin.

## I. Introduction

Whereas examples of entropic forces have been known for many years (e.g. forces arising from stretching rubber, from osmosis, and from stochastic processes in general), only recently has the relevance of the Brownian entropic force (BEF) first described by Neumann[1] come to be recognized. When determining the potential of mean force acting between pairs of molecules by means of molecular-dynamics simulation, it is frequently necessary to correct for the BEF and its associated entropy.[2,3] This same force has been used by Kharzeev[4] in a model describing the dissociation of quarkonium in terms of an "entropic self-destruction". In deriving this force[1] a Brownian particle, which at a given instant is separated from a fixed reference point by a scalar distance $r$, is viewed as having an entropy $S$ based on the number of configurations $W$ associated with $r$. Using the Boltzmann formula $S_B = k_B \ln W(r)$, the average entropic force $<f>$ is readily obtained from the thermodynamic relationship

$$<f> = T \, (dS_B/dr) = 2k_B T/r, \tag{1}$$

where $T$ is the absolute temperature; $k_B$ the Boltzmann constant; and $W(r) \propto 4\pi r^2$. $<f>$ is a repulsive force that acts so as to drive the particle away from the reference point.

Because the origin of the BEF remains obscure for many physical scientists and completely unrecognized by others[1,5] and because the usual derivations of Eq. (1) lack rigor, this work updates and upgrades the findings described in Ref. 1. The present approach also addresses the enumerability of $W(r)$ in the Boltzmann formula. The kinetic origin of the BEF is of particular interest because, to quote Weiner[6] in the context of polymer elasticity, "The role of entropy as a force potential is not directly physically intuitive. At first glance, it appears to describe a different category of force than that employed in Newton's equations of motion where, if forces are derived from a potential, the latter represents energy. If entropy is a measure of microscopic disorder, why does its gradient, multiplied by temperature, give rise to a macroscopic force?"

## II. Classical Canonical Ensemble and the Kinetic Origin of the Entropic Force

The ideal Brownian particle just described can be viewed as a particle of mass $m$, moving freely on the surface of a sphere of radius $r$. In spherical coordinates the appropriate Hamiltonian $H$ and partition function $Z$ are respectively

$$H = \frac{1}{2I}\left[ p_\theta^2 + \frac{p_\varphi^2}{\sin^2\theta} \right], \tag{2}$$

$$Z = \frac{1}{h^2} \int_{-\infty}^{+\infty} \int_{-\infty}^{+\infty} \int_0^{2\pi} \int_0^{\pi} \exp\left[-\frac{H}{k_B T}\right] d\theta \, d\varphi \, dp_\theta \, dp_\varphi = \frac{8\pi^2 k_B T I}{h^2}, \tag{3}$$

where $h$ = Planck's constant, $I = mr^2$, with spherical angles $\theta$ and $\varphi$, and their respective conjugate momenta $p_\theta$ and $p_\varphi$.[7] Using the standard relationships from statistical thermodynamics, one has $<f> = -(\partial A/\partial r)_T$, where the Helmholtz free energy is given by $A = -k_B T \ln Z$, resulting in $<f> = 2k_B T/r$.

The kinetic origin of the BEF may be understood following the approach of Weiner in discussing the retractive force observed when stretching a polymer molecule.[6] The thermodynamic definition of the Helmholtz energy, $A = U - TS$, leads to $(\partial A/\partial r)_T = (\partial U/\partial r)_T - T(\partial S/\partial r)_T$, where $U$ is the average energy of the particle, here assumed to be solely kinetic and independent of $r$. The entropic force can now be recast as, $<f> = -(\partial A/\partial r)_T = T(\partial S/\partial r)_T = k_B T (\partial \ln Z/\partial r)_T$ or from Eq. (3)

$$<f> = -\frac{k_B T}{Zh^2} \int_\Gamma \frac{1}{k_B T} \frac{\partial H}{\partial r} \exp\left[-\frac{H}{k_B T}\right] dq\, dp, \qquad (4)$$

where the single integral sign, $dq$, and $dp$ represent the multiple integral signs, differential coordinates, and differential momenta, respectively, shown in Eq. (3). From Eq. 2 one obtains $\partial H/\partial r = -2H/r$, which when substituted into Eq. 4 yields

$$<f> = \frac{1}{Zh^2} \int_\Gamma \frac{2H}{r} \exp\left[-\frac{H}{k_B T}\right] dq\, dp = \frac{2<H>}{r} = \frac{2k_B T}{r}. \qquad (5)$$

Because $H$ represents the rotational energy with two degrees of freedom, its ensemble-averaged value $<H>$ is $2(k_B T/2)$ or simply $k_B T$. Note that $<H> = U$. A kinetic origin for $<f>$ is now apparent in Eq. (5), which expresses this force as an ensemble average of the centrifugal force, $2H/r$. Note that this is the same force that causes centrifugal distortion in rotating diatomic molecules and must be taken into account when interpreting spectroscopic measurements involving transitions between rotational energy levels.

**III. The Gibbs Entropy**

The Gibbs entropy expression, which is more generally used than that of Boltzmann, is given by

$$S_G \propto -k_B \int_\Gamma g \ln g\, dq\, dp, \qquad (6)$$

where the normalized distribution function $g(q, p, r, T)$, which defines the probability density for finding the particle with position $q$ and momentum $p$ for particular values of $r$ and $T$, is given by $g = Z^{-1} \exp(-H/k_B T)$.[6] Substitution of $g$ into Eq. (6) results in

$$S_G = \frac{k_B}{h^2} \int_\Gamma \left(\ln Z + \frac{H}{k_B T}\right) g\, dq\, dp = k_B \ln Z + \frac{U}{T} = k_B \ln\left[Z \exp\left(\frac{U}{k_B T}\right)\right]. \qquad (7)$$

Because $U = k_B T$ in this particular case, we conclude that $S_G = k_B \ln(eZ)$ or

$$S_G = k_B \ln \frac{4\pi r^2}{\left(\Lambda/\sqrt{e}\right)^2}, \tag{8}$$

where $\Lambda = h/(2\pi m k_B T)^{1/2}$ and e is the exponential with e ≈ 2.7. $\Lambda$ is known as the de Broglie thermal wavelength and is a length of the order-of-magnitude of the wavelength of a particle having a kinetic energy equal to $k_B T$.

## IV. A Quantum Particle in a Box

If one thinks in terms of the particle-in-a box problem from elementary quantum mechanics, the square of half the thermal wavelength would approximate a region of occupancy for such a particle on a two-dimensional surface. The reason is that a linear dimension can accomodate an integral number of half wavelengths with a single antinode (maximum) in the wave function occurring in each individual region available for occupancy. The probability amplitude for a particle in a one-dimensional box is given by

$$\Psi(x) = A \sin\left(\frac{2\pi\sqrt{2mE}}{h} x\right) = \sqrt{\frac{2}{a}} \sin\left(\frac{n\pi x}{a}\right), \tag{9}$$

where $E$ is the particle's energy, $x$ the position of the particle, $a$ the size of the box, and $n$ the quantum number.[8] The expression in the center is that for a free particle, and that on the right reflects the boundary conditions dictated by the size of the box. $n$ is equal to the number of antinodal positions and is related to the quantized energy by $E = (nh)^2/(8ma^2)$. From Eq. (9), the wavelength is seen to be $\lambda = 2a/n$. For a two-dimensional box of surface area $a^2$, the number of antinodal points (hence available sites for occupation by the particle) is $n^2$ or $a^2/(\lambda/2)^2$. Thus the expression $(4\pi r^2)/(\Lambda/e^{1/2})^2$ can be regarded as the surface area of the sphere divided by the approximate thermal *half* wavelength squared resulting in the number of sites for occupancy on the sphere, i.e., $W$. Because $S_B = k_B \ln W$ is the definition for entropy usually attributed to Boltzmann, we have shown how the definition of entropy in Eq. (6) leads to an expression where the resulting $W$ is enumerable. As a point of interest, the entropy for an ideal-gas particle in three dimensions is

$$S_G = k_B \ln \frac{V}{\left(\Lambda/\sqrt{e}\right)^3}, \tag{10}$$

where $V$ is the volume of the container.[9] Equation (10) is of course totally consistent with the above discussion and is what one might expect for a particle in a three-dimensional lattice with the enumerable sites[10] available for occupancy being the antinodal points of standing waves.

## V. Conclusion

We have described an entropic force arising from Brownian motion using a traditional statistical-mechanical approach that includes momentum as well as configurational states by means of the classical Gibbs canonical partition function and thereby justified the use of Eq. (1). The kinetic origin of <*f*> was established through the ensemble averaging of the centrifugal force acting on the particle along *r*, the line segment connecting the particle with a fixed point or with a another particle serving as a reference point. In other words if two *ideal* particles, otherwise unencumbered, were joined by a thin, massless rod of length *r*, the average tension in the rod would be $2k_BT/r$. As noted, the kinetic origin of the BEF is also responsible for the appearance of this force in molecular-dynamics simulations and the need to correct for it when calculating potentials of mean force.[2,3]

Finally the role of the temperature must be emphasized for two reasons. (i) Classical statistical mechanics is valid only at temperatures sufficiently high to ensure the validity of the phase integrals involved. (ii) In order for a spherical surface to be divided up into regions each of approximate area $(\Lambda/2)^2$, the thermal wavelength must be small relative to the linear dimension; i.e., $\Lambda \ll r$. As $\Lambda$ varies inversely with the square root of the temperature, the model is invalid at low temperatures.

**ACKNOWLEDGEMENT**

I wish to thank Nico Roos for correspondence, Mehmet Süzen for calling my attention to Ref. 6, and Norma Sims Roche for encouragement and assistance in the preparation of this manuscript.

a) Electronic mail: richard.neumann@post.harvard.edu

**References**

[1] R. M. Neumann, "Entropic approach to Brownian Movement," Am. J. Phys. **48**(5), 354-357 (1980).
[2] C. Caleman, J. S. Hub, P. J. van Maaren, and D. van der Spoel, "Atomistic simulation of ion solvation in water explains surface preference of halides", Proc. Natl. Acad. Sci. U.S.A. **108**(17), 6838-6842 (2011).
[3] D. van der Spoel, E. Lindahl, B. Hess, and the GROMACS development team, *Gromacs User Manual version 4.6.5*, www.gromacs.org (2013), p. 151.
[4] D. E. Kharzeev, "Deconfinement as an entropic self-destruction: A solution for the quarkonium suppression puzzle?", Phys. Rev. D **90**(7), 074007 (2014).
[5] Polymer physics texts do not include the BEF when calculating the entropic force that exists between the ends of a polymer chain. e.g., see M. Rubinstein and R. H. Colby, *Polymer Physics* (Oxford University Press, Oxford, 2003). Ref. 1 describes some paradoxical consequences resulting from the omission of this force.
[6] J. H. Weiner, "Entropic versus kinetic viewpoints in rubber elasticity", Am. J. Phys. **55**(8), 746-749 (1987).
[7] T. L. Hill, *An Introduction to Statistical Thermodynamics* (Addison-Wesley, Reading, MA, 1962), pp. 154-155.
[8] D. A. McQuarrie and J. D. Simon, *Physical Chemistry a Molecular Approach* (University Science Books, Sausalito, CA, 1997), pp. 80-85.

[9] J. E. Mayer and M. G. Mayer, *Statistical Mechanics*, Second Edition (John Wiley & Sons, New York, 1977), p. 171.
[10] W. Greiner, L. Neise, and H. Stöcker, *Thermodynamics and Statistical Mechanics*, Springer Verlag New York, Inc., New York, 1995), p. 140.